\documentclass[12pt,a4paper]{article}
\pdfoutput=1
\usepackage{amsmath}
\usepackage{amsfonts}
\usepackage{amssymb}
\usepackage{mathtools}
\usepackage{bbm}
\usepackage{verbatim}
\usepackage{dsfont}
\usepackage{booktabs}
\usepackage{slashed}
\usepackage{comment}
\usepackage{epsfig}
\usepackage{color}
\usepackage[table]{xcolor}
\usepackage{graphicx}
\usepackage{caption}
\usepackage{subfigure}
\usepackage{float}
\usepackage{overpic}
\usepackage{units}

\usepackage{enumerate}
\usepackage{hhline}
\usepackage{multirow}

\usepackage{cite}
\usepackage{xspace}
\usepackage{setspace}

\usepackage[pdftitle={},
  pdfauthor={},
  pdfsubject={},
  bookmarksopen, bookmarksnumbered, bookmarksopenlevel=2, colorlinks=false, linkcolor=blue, citecolor=blue, urlcolor=blue]{hyperref}

\newcommand{\beqa}{\begin{eqnarray}}
\newcommand{\eeqa}{\end{eqnarray}}
\newcommand{\be}{\begin{equation}}
\newcommand{\ee}{\end{equation}}
\newcommand{\ba}{\begin{array}} 
\newcommand{\ea}{\end{array}}

\begin{document}

\thispagestyle{empty}

\begin{flushright}
DESY 19-049\\
\end{flushright}
\vskip .8 cm
\begin{center}
	{\Large {\bf Proton decay in flux compactifications
}}\\[12pt]

\bigskip
\bigskip 
{
{\bf{Wilfried Buchmuller$^a$}\footnote{E-mail:
    wilfried.buchmueller@desy.de}} and
{\bf{Ketan M. Patel$^b$}\footnote{E-mail: kmpatel@prl.res.in}}
\bigskip}\\[0pt]
\vspace{0.23cm}
{\it $^a$ Deutsches Elektronen-Synchrotron DESY, 22607 Hamburg, Germany \\ }

{\it $^b$ Physical Research Laboratory, Navarangpura, Ahmedabad  380 009, India \\ \vspace{0.2cm}
}
\bigskip
\end{center}

\begin{abstract}
\noindent
We study proton decay in a six-dimensional orbifold GUT model with gauge
group $SO(10)\times U(1)_A$. Magnetic $U(1)_A$ flux in the compact
dimensions determines the multiplicity of quark-lepton generations,
and it also breaks supersymmetry by giving universal GUT scale masses to scalar quarks and
leptons. The model can successfully account for quark and lepton masses
and mixings.  Our analysis of proton decay
leads to the conclusion that the proton lifetime must be close to the
current experimental lower bound. Moreover, we find that the branching
ratios for the decay channels $p \rightarrow e^+\pi^0$ and
$p\rightarrow \mu^+\pi^0$ are of similar size, in fact the latter one can 
even be dominant. This is due to flavour non-diagonal couplings of heavy
vector bosons together with large off-diagonal
Higgs couplings, which appears to be a generic feature of flux compactifications. 
\end{abstract}

\newpage 
\setcounter{page}{2}
\setcounter{footnote}{0}

{\renewcommand{\baselinestretch}{1.5}
\section{Introduction}
\label{sec:introduction}

Proton decay is a striking prediction of Grand Unified Theories (GUTs)
which predict interactions violating baryon number ($B$) and lepton
number ($L$)
\cite{Pati:1973uk,Georgi:1974sy}. In non-supersymmetric theories proton decay is caused by
the exchange of heavy vector bosons, leading to $B$+$L$ violating
dimension-six operators with dominant decay mode $p\rightarrow
e^+\pi^0$. In supersymmetric theories the exchange of colour-triplet
scalars can lead to dangerous $B$+$L$ violating dimension-five
operators \cite{Weinberg:1981wj,Sakai:1981pk} 
and $p\rightarrow \bar{\nu}K^+$ as dominant decay mode. Consistency with the observed proton
lifetime then requires that the colour-triplet partners of Higgs bosons are very heavy \cite{Hisano:1992jj}.
It is remarkable that these dangerous operators are
generically absent in higher-dimensional orbifold GUTs \cite{Altarelli:2001qj,
Hall:2001pg,Hebecker:2001wq}
where, on the other hand, dimension-six operators can be enhanced by
Kaluza-Klein towers of heavy vector bosons.

The discovery of proton decay would not only strongly support the idea
of grand unification, the measurement of the proton lifetime and its
branching ratios would also provide valuable information on the mass
scale of unification and on the flavour structure of the
theory. Moreover, the pattern of proton decays would help to
distinguish between the different GUT groups, including
the Pati-Salam group
\cite{Pati:1973uk}, Georgi-Glashow $SU(5)$ \cite{Georgi:1974sy},
$SO(10)$ \cite{Georgi:1974my,Fritzsch:1974nn} or flipped $SU(5)$
\cite{Barr:1981qv,Derendinger:1983aj}.

Proton decay has already been studied for a large variety
of GUT models (for reviews see, for
example, \cite{Langacker:1980js,Nath:2006ut,Raby:2017ucc,Tanabashi:2018oca}). 
Here we are particularly interested in predictions for proton decay
in orbifold GUT models. During the past years a number of analyses have been carried
out for five-dimensional (5D) models (see, for example, \cite{Altarelli:2001qj,
Hall:2001pg,Hebecker:2001wq,Dermisek:2001hp,
Hebecker:2002rc,Shafi:2002ii,Kim:2002im,Alciati:2005ur}) and also for
six-dimensional (6D) models \cite{Buchmuller:2004eg,deAnda:2018yfp,deAnda:2018oik}.
Recently, a new class of magnetized 6D orbifold GUT models with broken
supersymmetry has been studied \cite{Buchmuller:2015jna}. 
The magnetic flux in the extra dimensions plays a twofold
role, it generates the multiplicity of quark-lepton generations and it
also breaks supersymmetry \cite{Bachas:1995ik} around the GUT scale. The effective
low-energy theory contains two Higgs doublets and possibly light
higgsinos, whereas scalar quarks and leptons have GUT scale masses.
The theory can account for quark and lepton masses and
mixings \cite{Buchmuller:2017vho,Buchmuller:2017vut}, and it can be consistently matched to a supersymmetric
theory at the compactification scale, which approximately equals the
GUT scale \cite{Lee:2015uza,Bagnaschi:2015pwa,Mummidi:2018nph}. Light higgsinos can be made consistent with
constraints from direct detection by mixing with an additional singlet
\cite{Mummidi:2018myd}. 

In this paper we study proton decay in magnetized orbifold GUTs. The
decay rate crucially depends on the compactification scale $M_c$ which is
constrained by the requirement of gauge coupling unification at some
cutoff-scale $\Lambda$ of the higher-dimensional theory \cite{Dienes:1998vg}.
Since the full ultraviolet completion of the 4D effective theory is not known,
the compactification scale can only be estimated. This is in contrast to 4D GUTs where 
the masses of heavy vector bosons can be precisely computed for
theories in which the Standard Model gauge couplings unify. In a
detailed analysis of 5D orbifold GUTs the authors find an uncertainty
of about two orders of magnitude for $M_c$ \cite{Alciati:2005ur}. 
Recently, a comparison of 4D GUTs and 5D orbifold GUTs has been given
in \cite{Pokorski:2019ete}.

It is well known that branching ratios in proton decay strongly depend
on the flavour structure of the theory. We find comparable branching
ratios for $p\rightarrow e^+\pi^0$ and
$p\rightarrow \mu^+\pi^0$. As we shall see, this is partly due to
non-diagonal coupling of heavy vector bosons to flavour eigenstates,
which is caused by non-trivial overlap integrals of fermion and vector
boson mode functions. A large branching ratio to $\mu^+\pi^0$, due to large mixing among
charged leptons, has previously been discussed in the context of a flipped $SU(5)$
model \cite{Ellis:2002vk}. It is intriguing, that the Super-Kamiokande
collaboration has observed two candidates for $p\rightarrow 
\mu^+\pi^0$  which, however, are consistent with the expected number of
background events \cite{Miura:2016krn}.

The paper is organized as follows. In Section~2 we briefly recall the
main features of our model and we discuss the constraints on the
compactification scale based on gauge coupling unification. Section~3
deals with the various contributions to proton decay and determines
lifetime and branching ratios into different final states. Our
conclusions are given in Section~4. In Appendix~A the relevant overlap
integrals of mode functions are given, and Appendix~B contains the
unitary matrices that connect weak and mass eigenstates for two fits
of the model parameters to quark and lepton masses and mixings.

\section{GUT model and gauge coupling unification}
\label{sec:model}

The considered 6D GUT model has been described in detail in
\cite{Buchmuller:2015jna,Buchmuller:2017vho,Buchmuller:2017vut}. The
bulk gauge group is $SO(10)\times U(1)_A$. The 6D theory is compactified
on the orbifold $T^2/\mathbb{Z}_2$, and the GUT gauge group is broken to the Standard
Model gauge group by two Wilson-lines (see Fig.~\ref{fig1}). At the
fixed point $\zeta_{\rm I}$ the bulk $SO(10)$ symmetry is left unbroken
whereas at the fixed points $\zeta_{\rm PS}$, $\zeta_{\rm GG}$ and $\zeta_{\rm fl}$
it is broken to three different subgroups, respectively\cite{Asaka:2001eh,Hall:2001xr},
\begin{align}
G_\mathrm{PS} &= SU(4)\times SU(2)_{\mathrm{L}}\times SU(2)_{\mathrm{R}} \nonumber\\
&\supset SU(3)\times U(1)_{B-L}\times SU(2)_{\mathrm{L}}\times
U(1)_{\rm R} \ ,\\
G_\mathrm{GG} &= SU(5)\times U(1)_X \nonumber\\
&\supset SU(3)\times SU(2)_{\mathrm{L}}\times U(1)_Y \times U(1)_X\ ,\\
G_\mathrm{fl} &= SU(5)' \times U(1)_{X'} \nonumber\\
&\supset SU(3)\times SU(2)_{\mathrm{L}}\times U(1)_Z \times U(1)_{X'}\ .
\end{align} 
Clearly, the Standard Model gauge group with an additional $U(1)$
factor can be obtained as intersection of  the subgroups at two
different fixed points. 

The relations between the generators of the different $U(1)$ factors
are easily obtained by considering the decomposition of the $SO(10)$
$16$-plet, in standard notation $16 \supset d^c,
l,q,u^c,e^c,n^c$, at the different fixed points. At $\zeta_{\rm GG}$, with
$U(1)_Y \times U(1)_X$, one has $16 \supset 5^*_3 + 10_{-1} + 1_{-5}$, with $5^*_3 \supset (3^*,1)_{1/3,3} +
(1,2)_{-1/2,3} \sim d^c + l$, $10_{-1} \supset (3,2)_{1/6,-1} +
(3^*,1)_{-2/3,-1} + (1,1)_{1,-1}  \sim q + u^c + e^c$ and $1_{-5} =
(1,1)_{0,-5} \sim n^c$.
Correspondingly, at $\zeta_{\rm PS}$, with
$U(1)_{B-L} \times U(1)_R$, the decomposition reads 
$16 \supset (4,2,1) + (4^*,1,2)$, with $(4^*,1,2) \supset
(3^*,1)_{-1/3,1/2} + (3^*,1)_{-1/3,-1/2} + (1,1)_{1,1/2} + (1,1)_{1,-1/2}
\sim d^c + u^c + e^c + n^c$. Comparison with the decomposition at
$\zeta_{\rm GG}$ yields the relations
\begin{equation}\label{U1sPS}
B-L = \frac{4}{5}Y - \frac{1}{5}X\ , \quad 
I_{3R} = \frac{3}{5}Y + \frac{1}{10} X \ .
\end{equation}
At $\zeta_{\rm fl}$, with $U(1)_{Z} \times U(1)_{X'}$,
the decomposition is the same as at $\zeta_{\rm GG}$, with $Z$ and
$X'$ taking the role of $Y$ and $X$, respectively.
Flipped $SU(5)$ is obtained from Georgi-Glashow $SU(5)$ by exchanging
$d^c$ and $u^c$, and $n^c$ and $e^c$. This implies
$16 \supset 5^{*'}_3 + 10'_{-1} + 1'_{-5}$, with $5^{*'}_3 \supset (3^*,1)_{1/3,3} +
(1,2)_{-1/2,3} \sim u^c + l$, $10_{-1} \supset (3,2)_{1/6,-1} +
(3^*,1)_{-2/3,-1} + (1,1)_{1,-1}  \sim q + d^c + n^c$ and $1_{-5} =
(1,1)_{0,-5} \sim e^c$. From comparison with the decomposition at
$\zeta_{\rm GG}$ one obtains
\begin{equation}\label{U1sfl}
Z = -\frac{1}{5}(Y+X)\ , \quad
X' = -\frac{24}{5}Y + \frac{1}{5} X\ .
\end{equation}

\begin{figure}
\centering 
\begin{overpic}[scale = 0.3, tics=10]{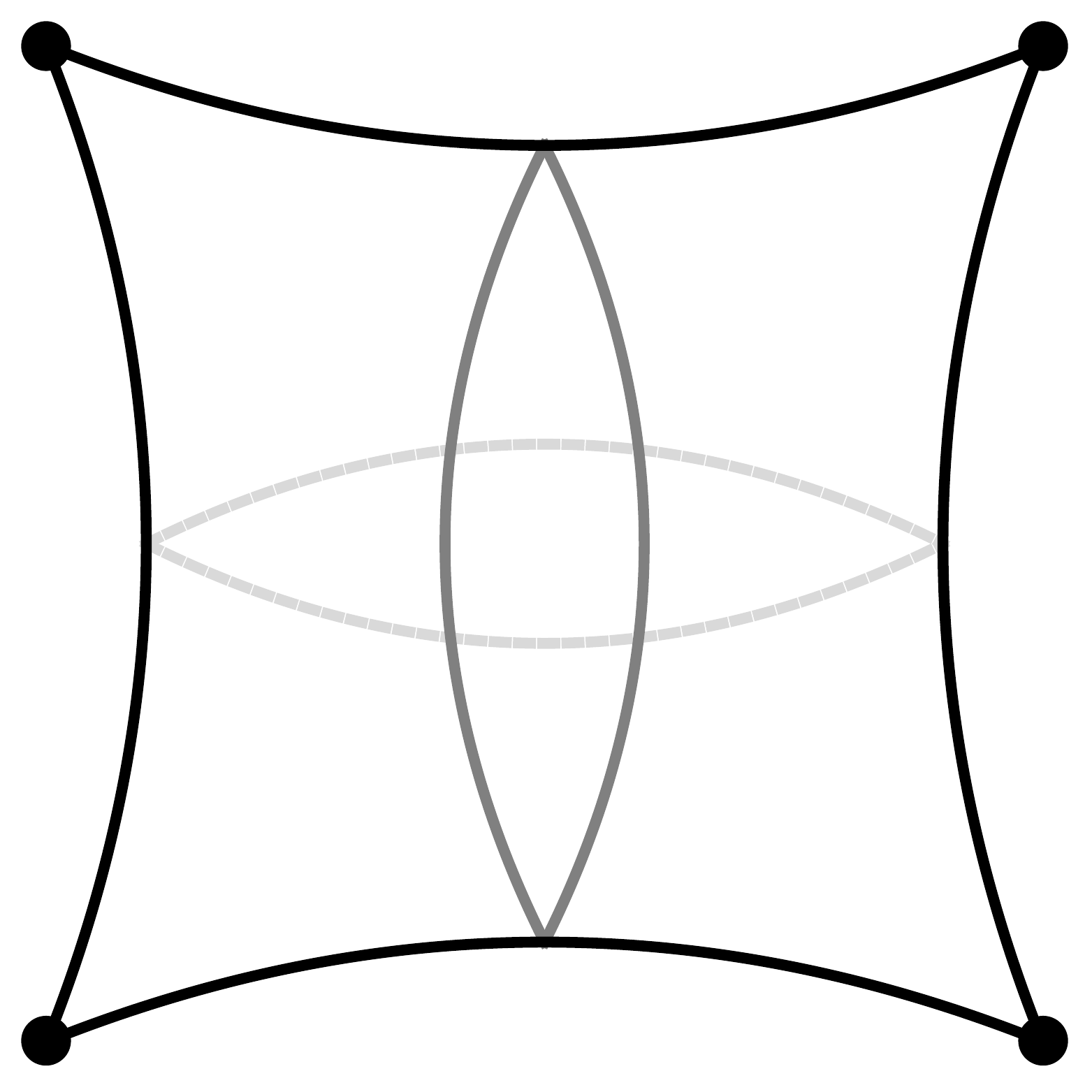}
	\put(-7, 2){$\zeta_{\mathrm{I}}$}
	\put(100, 2){$\zeta_{\mathrm{PS}}$}
	\put(100, 97){$\zeta_{\mathrm{fl}}$}
	\put(-12, 97){$\zeta_{\mathrm{GG}}$}
\end{overpic}
\caption{Orbifold $T^2/{\mathbb Z_2}$  with two Wilson lines and the
fixed points $\zeta_{\rm I}$, $\zeta_{\rm PS}$, $\zeta_{\rm GG}$, and $\zeta_{\rm fl}$.}
\label{fig1}
\end{figure}
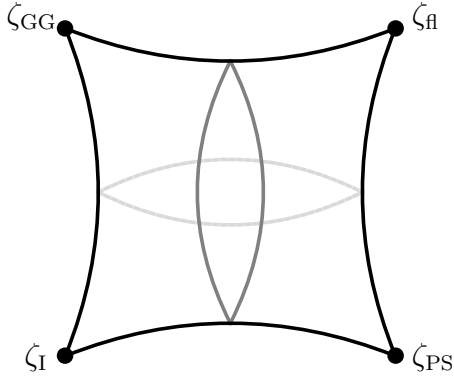
The 6D theory has six 16-plets two of which, $\psi$ and $\chi$, contain the three
quark-lepton generations as zero-modes.
In addition, they yield as split multiplets a fourth set of
quark and lepton $SU(2)_\text{L}$ singlets, $u^c$, $d^c$, $n^c$, $e^c$. Two
16-plets, $\psi^c$ and $\chi^c$, contain the charge-conjugate singlets 
$u$, $d$, $n$, $e$. Each of them decouples via a GUT scale mass term one linear
combination of the four sets of ${u^c}$, ${d^c}$, ${n^c}$ and ${e^c}$, respectively.
The three orthogonal linear combinations remain
in the low-energy spectrum. 
Furthermore, there are two 16-plets, $\Psi$
and $\Psi^c$, yielding singlets for spontaneous $B-L$ breaking, and 
eight 10-plets required by 6D anomaly cancellation, which contain two
Higgs doublets, $H_u$ and $H_d$, and further vector-like split
multiplets that acquire GUT scale masses. The vector-like split
multiplets are crucial to obtain realistic mass matrices for the light
quarks and leptons, as discussed in \cite{Buchmuller:2017vho,Buchmuller:2017vut}.
However, they have no effect on proton decay amplitudes and we can
therefore ignore them in the following. The 16-plets $\psi$ and $\chi$
have nonvanishing $U(1)_A$ charge which determines the multiplicity of
their zero-modes in the magnetic flux background. Explicitly, the mode expansions
read
\begin{equation} \label{decomposition}
\begin{split}
\psi &= \sum_{i'=1,2} \left[ q_{i'} \psi_{-+}^{(i')}\, + \, l_{i'} \psi_{--}^{(i')}\, + \, (d^c_{i'}+n^c_{i'}) \psi_{+-}^{(i')} \right] + \sum_{\alpha'=1,2,3} (u^c_{\alpha'}+e^c_{\alpha'}) \psi_{++}^{(\alpha')}\,,  \\
\chi &= q_3 \chi_{--}^{(1)}\, + \, l_3 \chi_{-+}^{(1)}\, + \,
(u^c_4+e^c_4) \chi_{+-}^{(1)}\, + \, \sum_{i'=1,2}(d^c_{i'+2}
+n^c_{i'+2}) \chi_{++}^{(i')} \,,
\\
\psi^c &= u+e\,,  \quad
\chi^c = d+n\,,\quad \Psi = D^c + N^c\,,\quad \Psi^c = D + N\,.
\end{split}
\end{equation}
Since the GUT group $SO(10)$ is broken by two Wilson-lines, the
spectrum of fermion zero-modes is determined by the choice of two
parities, for instance at the Pati-Salam (PS) fixed point and the
Georgie-Glashow (GG) fixed point. Since the parities are associated with
matrices that do not commute with $SO(10)$, the parities of the
different Standard Model fields contained in the 16-plets are
in general different. The choice in \cite{Buchmuller:2017vut} leads to
the decompositions given in Eqs.~\eqref{decomposition}. The
subscripts of the 4D fields, $i'$ and $\alpha'$, label the
degeneracy of the corresponding zero-modes. In addition to quarks and
leptons the low-energy theory contains two light Higgs doublets, and
possibly a light $SU(2)_\text{L}$ doublet pair of higgsinos. 

The proton lifetime crucially depends on the compactification scale $M_c$,
which is related to the scale of unification in
higher-dimensional GUTs. In orbifold GUTs, where the GUT symmetry is
broken at orbifold fixed points, gauge couplings unify only
approximately at the compactification scale, and corrections
due to massive vector-like split multiplets, brane kinetic terms and the field content of
the higher-dimensional theory have to be taken into account to achieve
unification of couplings at some cut-off-scale $\Lambda > M_c$. In the
following we study the consistency between proton decay and constraints from gauge
coupling unification.

\subsection{Brane kinetic terms}

The 4D gauge couplings receive contributions from the 6D bulk gauge
coupling and from brane kinetic terms,
\begin{equation}
\frac{1}{g_i^2(M_c)} = \frac{V_2}{g^2_{6D}(M_c)} +
\sum_{p={\rm I}, {\rm PS},{\rm GG}, {\rm fl}} 
\frac{1}{g^2_i(M_c)}\Big|_p \ .
\end{equation}
Here $V_2$ is the volume of the orbifold, and the brane kinetic terms
take the form
\be \label{kin_brane}
{\cal L}_{\rm 4D}^{\rm GK} = 
 -\frac{1}{4}\sum_p \hat{F}^2 |_p \equiv -\frac{1}{4}\sum_{p,i} G^p_i F^2_{p,i}
\ ,
\ee
where $G^p_i$ are constants and $F^2_{p,i} =  F^a_{\mu\nu}
F^{a\,\mu \nu}|_{p,i}$ is a field strength squared for a factor of
the unbroken gauge group at fixed point $p$. For our $SO(10)$
model one has
\be \label{F_I}
F^2|_{\rm I} = G^{\rm I}\, F_{45}^2\ ,
\ee
\be \label{F_PS}
F^2|_{\rm PS} = G^{\rm PS}_1\, F_{(15,1,1)}^2 + G^{\rm PS}_2\, F_{(1,3,1)}^2 + G^{\rm PS}_3\, F_{(1,1,3)}^2\,,\ee
\be \label{F_GG_fl}
F^2|_q= G^q_1\, F_{(24,0)}^2 + G^q_2\, F_{(1,0)}^2  \,,
\ee
with $q={\rm GG}, {\rm fl}$. 
At each fixed point the unbroken gauge group contains two $U(1)$
factors: $U(1)_Y \times U(1)_X$ at $\zeta_{\rm I}$ and $\zeta_{\rm GG}$,
$U(1)_Z \times U(1)_{X'}$ at $\zeta_{\rm fl}$, and $U(1)_{B-L} \times
U(1)_{R}$ at $\zeta_{\rm PS}$. Using Eqs.~\eqref{U1sPS} and \eqref{U1sfl}, the
$U(1)$ gauge kinetic terms for $B-L$, $I_{3R}$, $Z$ and $X'$ can be
expressed in terms of kinetic terms for $U(1)_Y$ and $U(1)_X$. In this
way one obtains for the gauge kinetic terms of the unbroken 4D gauge
group $SU(3)\times SU(2)\times U(1)_Y \times U(1)_X$:
\begin{align} \label{kin_brane_res}
-4 {\cal L}_{\rm 4D}^{\rm GK}  &\supset \left(\hat{G} + G^{\rm
    PS}_1+ G^{\rm GG}_1 + G^{\rm fl}_1 \right)F_{(8,1,0,0)}^2 
+ \left(\hat{G} + G^{\rm PS}_2+ G^{\rm GG}_1 + G^{\rm fl}_1 \right) F_{(1,3,0,0)}^2  \nonumber\\
& +  \left(\frac{3}{5}\hat{G} + \frac{6}{25} G^{\rm PS}_1+ \frac{9}{25} G^{\rm PS}_3 + \frac{3}{5} G^{\rm GG}_1+ \frac{3}{125} G^{\rm fl}_1 + \frac{72}{125} G^{\rm fl}_2 \right)\, F_Y^2\,  \nonumber\\
& +  \left(\frac{1}{40}\hat{G} + \frac{3}{200} G^{\rm PS}_1+
  \frac{1}{100} G^{\rm PS}_3 + \frac{1}{40} G^{\rm GG}_2+
  \frac{3}{125} G^{\rm fl}_1 + \frac{1}{1000} G^{\rm fl}_2 \right)\,
F_X^2\, ,
\end{align}
with $\hat{G} = V_2/g^2_{6D} + G^{\rm I}$.
It is convenient to normalize the SM and $U(1)_X$ gauge couplings
to the colour $SU(3)$ coupling, which yields
\begin{align} \label{SM_gauge}
-4 {\cal L}_{\rm 4D}^{\rm GK}  \supset  &\ G F_{(8,1,0,0)}^2 + \left(G
  + \Delta_2^{\rm PS} \right) F_{(1,3,0,0)}^2 
+ \left(G + \frac{3}{5} \Delta_1^{\rm PS} + \frac{24}{25} \Delta^{\rm fl} \right) \frac{3}{5}F_Y^2 \nonumber \\
&+ \left(G+\frac{2}{5} \Delta^{\rm PS}_1 + \Delta^{\rm GG} +
  \frac{1}{25}\Delta^{\rm fl} \right) \frac{1}{40}F_X^2\,, 
\end{align}
where 
\begin{equation}
\begin{split}
G &= \hat{G} + G^{\rm PS}_1+G^{\rm GG}_1+G^{\rm fl}_1, \quad
\Delta^{\rm PS}_2 = G^{\rm PS}_2-G^{\rm PS}_1\ , \\ 
\Delta^{\rm PS}_1 &= G^{\rm PS}_3-G^{\rm PS}_1, \quad 
\Delta^{\rm fl} = G^{\rm fl}_2-G^{\rm  fl}_1, \quad \Delta^{\rm GG} =
G^{\rm GG}_2-G^{\rm GG}_1\ .
\end{split}
\end{equation} 
At the compactification scale, one then obtains the following relations between the SM gauge couplings:
\begin{equation} \label{gauge_uni}
\begin{split}
\frac{1}{\alpha_2}-\frac{1}{\alpha_3} &= 4\pi \Delta^{\rm PS}_2\, \equiv \delta_{23}\,, \\
\frac{1}{\alpha_1}-\frac{1}{\alpha_2} &= 4\pi\left(\frac{3}{5}
  \Delta^{\rm PS}_1 + \frac{24}{25} \Delta^{\rm fl} -\Delta^{\rm
    PS}_2\right)\, \equiv \delta_{12}\,,
\end{split}
\end{equation}
where $\alpha_1 = \frac{5}{3} \alpha_Y$. We do not discuss gauge
coupling unification for $U(1)_X$ since we assume that this $U(1)$
symmetry is spontaneously broken close to the compactification scale.

\subsection{The scale of compactification}

The conditions for gauge coupling unification given in
Eqs.~(\ref{gauge_uni}) need to be satisfied at the compactification scale
$M_c$. Since the two terms $\delta_{23}$ and $\delta_{12}$ are linear
combinations of five independent brane kinetic terms it is certainly
possible to satisfy these equations. However, it is not clear
whether this can be achieved with reasonable values for the brane
kinetic terms. Here we neglect further heavy threshold corrections and
higher-order running effects.

Let us first determine the phenomenologically required size of the
correction terms $\delta_{23}$ and $\delta_{12}$ in the two cases of a
pure two-Higgs-doublet model
(THDM) (case A) and a THDM with higgsinos of mass $1$~TeV (case B). The
three running couplings of the Standard Model gauge group and the
parameters $\delta_{23}$ and $\delta_{12}$ are shown for the two cases
in Fig.~\ref{fig2} and Fig.~\ref{fig3}, respectively, where we have
used one-loop renormalization group equations.
%
\begin{figure}[ht!]
\centering
\subfigure{\includegraphics[width=0.47\textwidth]{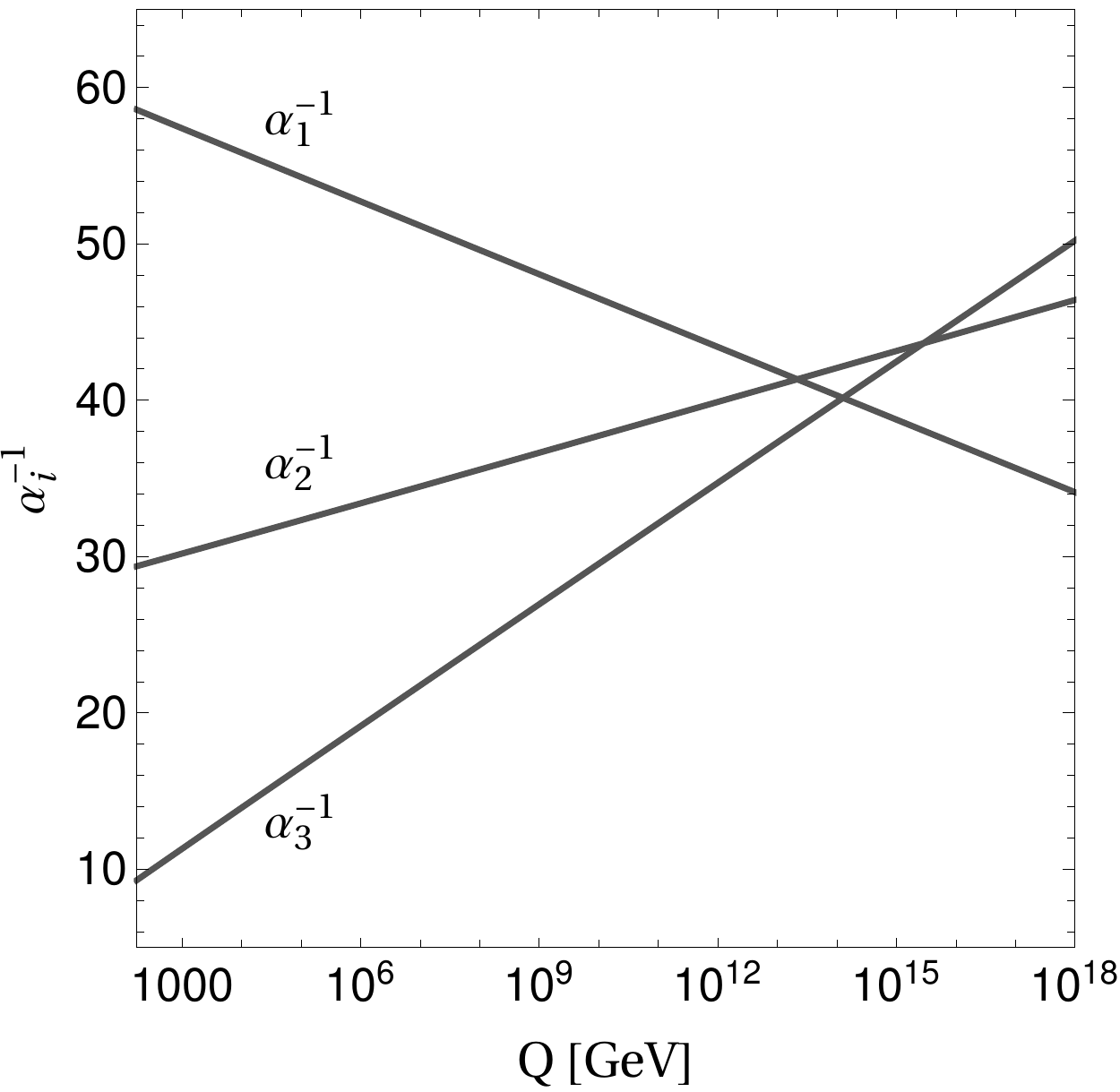}} \hspace*{0.0cm}
\subfigure{\includegraphics[width=0.47\textwidth]{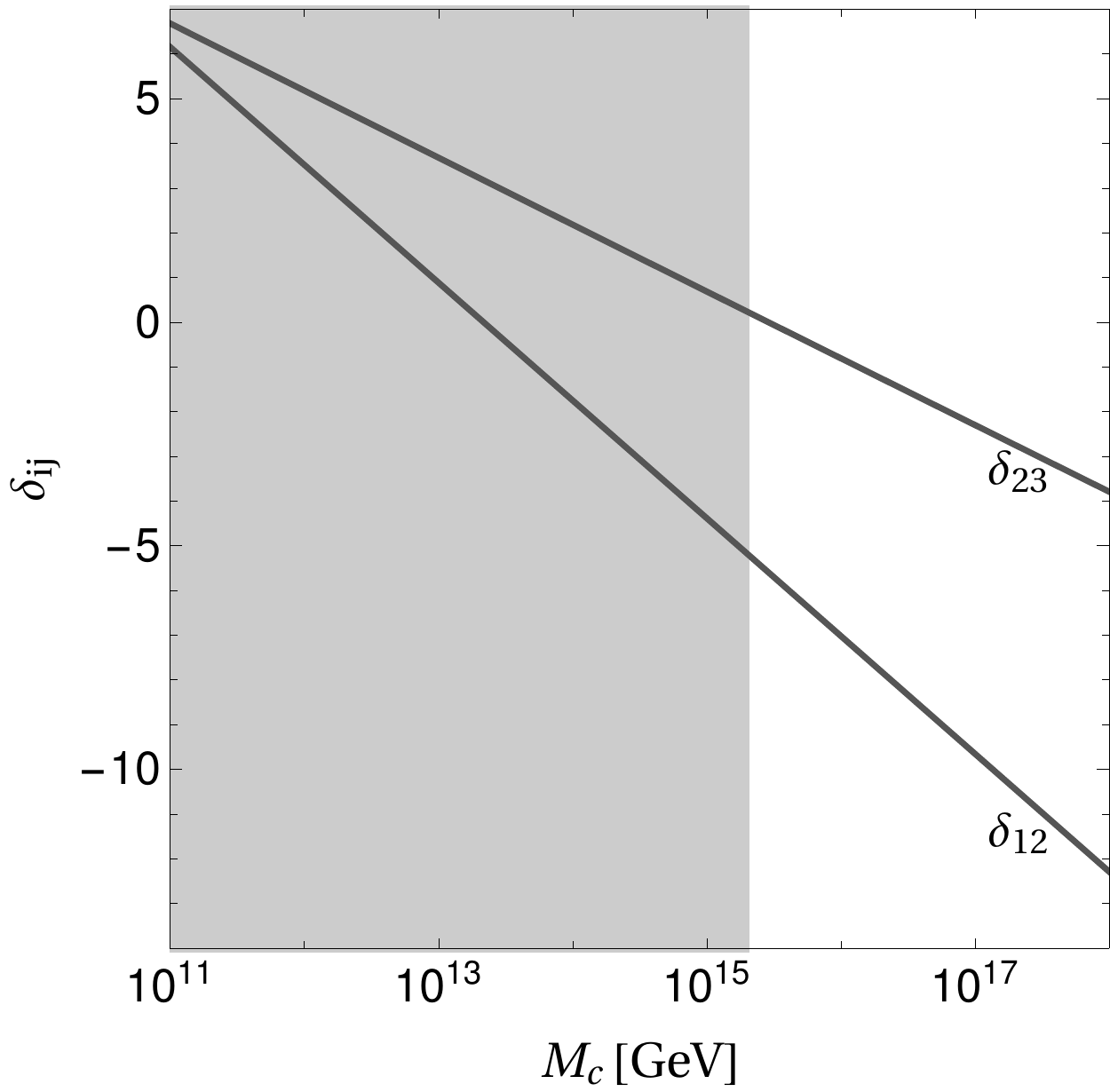}}\\
\caption{Left panel: Running of $\alpha_i^{-1}$ with respect to
  renormalization scale $Q$ for case A. Right panel: The 
  values of $\delta_{23}$ and $\delta_{12}$ required by the conditions given in
   Eq.~\eqref{gauge_uni}. The gray region is excluded by the current
   lower bound on the proton lifetime.}
\label{fig2}
\end{figure}
\begin{figure}[ht!]
\centering
\subfigure{\includegraphics[width=0.47\textwidth]{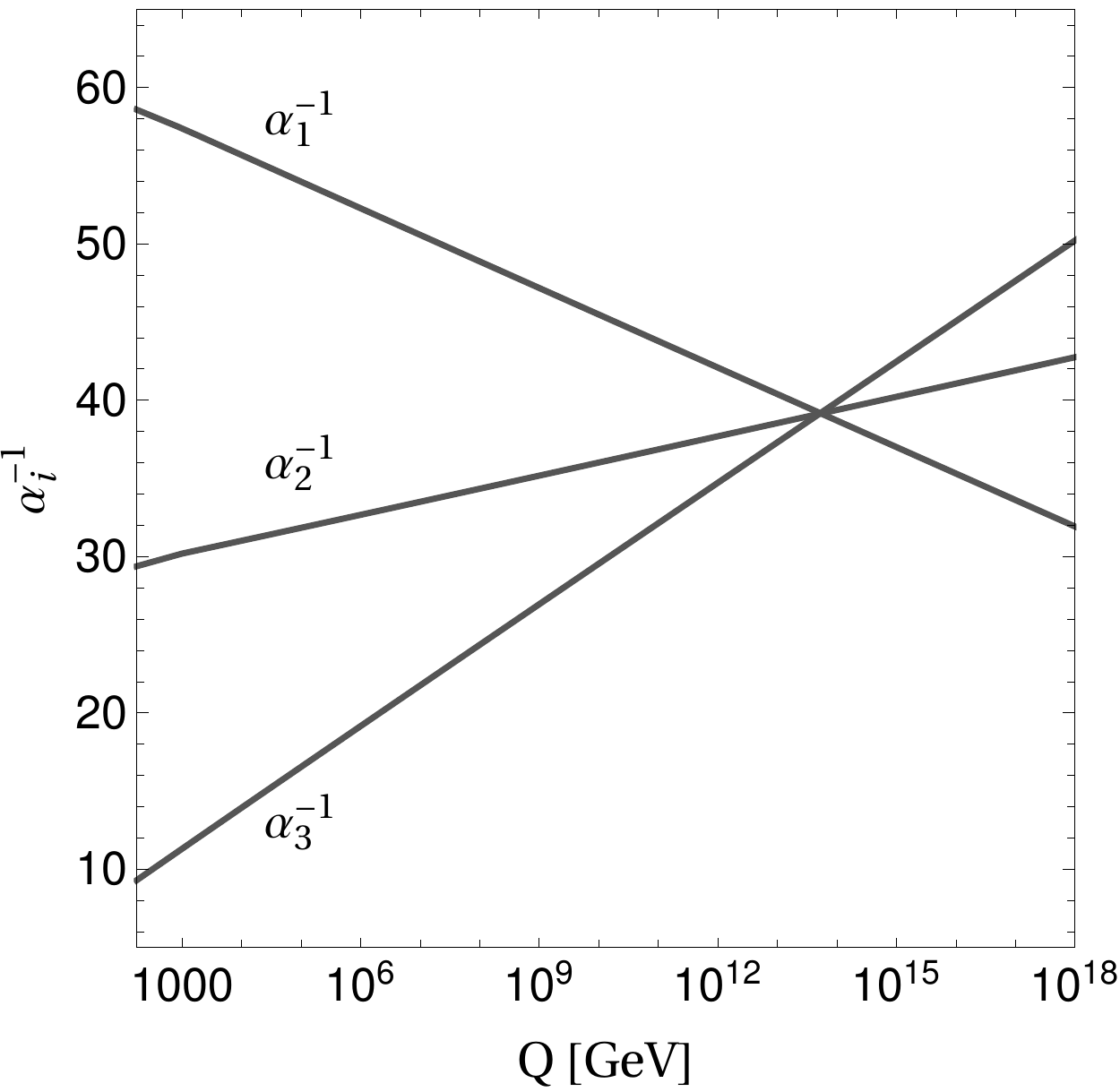}} \hspace*{0.1cm}
\subfigure{\includegraphics[width=0.47\textwidth]{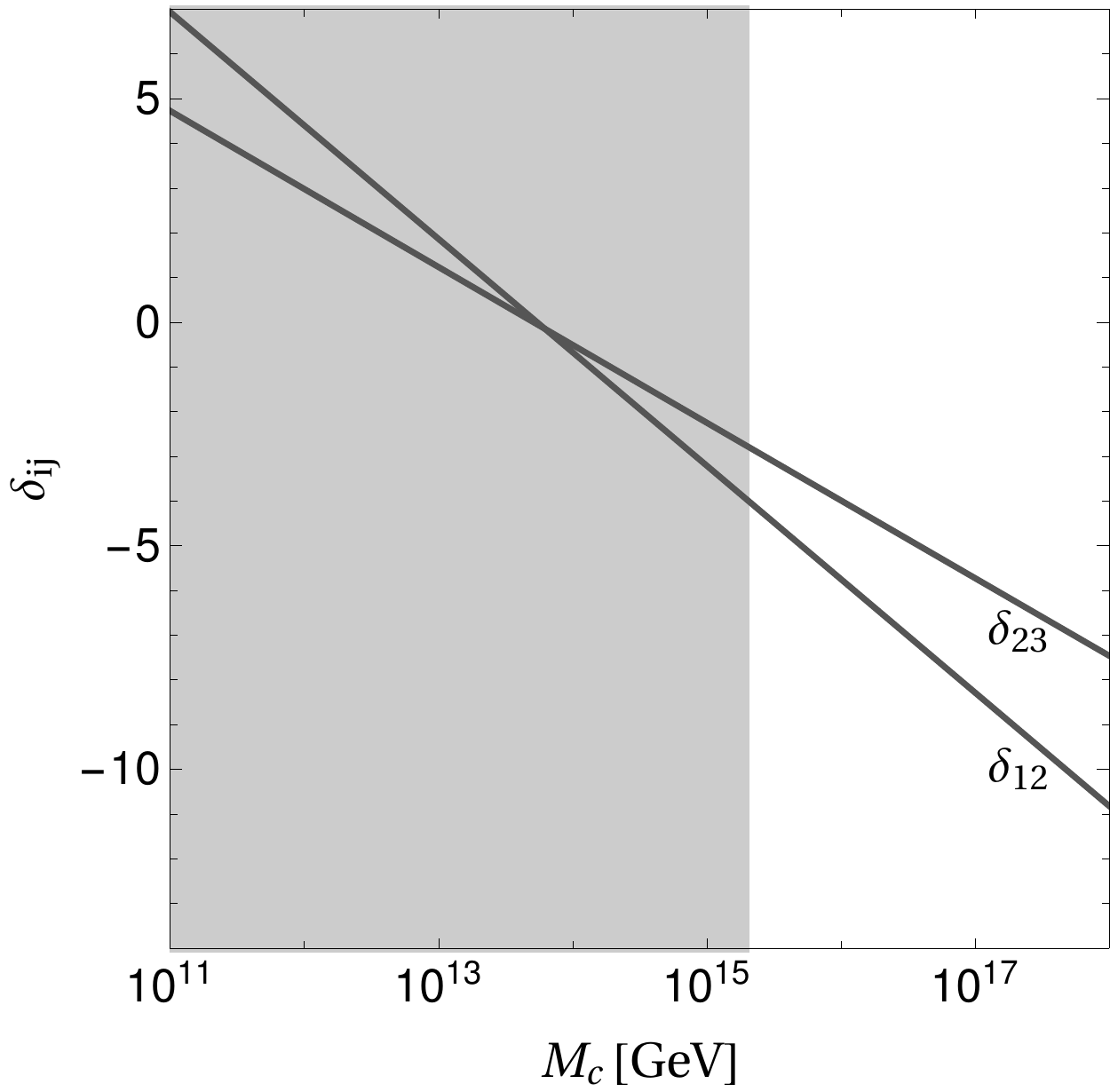}}\\
\caption{Same as in Fig. \ref{fig2}, but for the case B.}
\label{fig3}
\end{figure}
With light higgsinos (Fig.~\ref{fig3}), one
obtains a rather accurate gauge coupling unification without
correction terms at $Q \sim 10^{14}~\text{GeV}$. As a consequence, the
difference $|\delta_{23}-\delta_{12}|$ is smaller in case~B than in
case~A. In the right panel of Fig.~2 and Fig.~3 the gray region ($M_c
< 10^{15}~\mathrm{GeV}$) is excluded by the current lower bound on the
proton lifetime (see Section~3).

In higher-dimensional theories gauge couplings receive power-law
quantum corrections for scales above the compactification scale,
$\Lambda > M_c$. In six dimensions one has at one-loop order \cite{Dienes:1998vg},
\begin{equation}
\delta\left(\frac{1}{g_i^2(\Lambda)}\right) = - \frac{b\pi}{16\pi^2} \left(\frac{\Lambda}{M_c}\right)^2\ , 
\end{equation} 
where $b$ is the one-loop coefficient of the $\beta$-function. 
For the gauge group
$SO(10)$ with $\mathcal{N}=2$ supersymmetry, eight 10-plets and six 16-plets,
which corresponds to the model in \cite{Buchmuller:2017vut},
one obtains\footnote{Here we have used $b=-\frac{11}{3}C_A + \frac{2}{3}T_R n_f
  + \frac{1}{2}T_R n_s$, where $n_f$ and $n_s$ are the number of Weyl
  fermions and complex scalars, respectively. Choosing $T_\text{10} = \frac{1}{2}$, one has
  $C_A = 4$ and $T_\text{16}=1$ for the gauge group $SO(10)$ \cite{Slansky:1981yr}.}
$b=12$. Due to the large number of bulk matter fields one reaches the
strong coupling regime, $1/g_i^2 \approx 0$,
already close to the compactification scale, at $\Lambda/M_c \simeq 3.7$. 
Assuming now gauge coupling unification at strong coupling at a cutoff $\Lambda > M_c$\cite{Nomura:2001tn},
one can estimate the size of brane kinetic terms due to quantum
corrections, 
\begin{equation}
G^p_i = \frac{b_{p,i}}{8\pi^2} \ln\left(\frac{\Lambda}{M_c}\right)\ .
\end{equation}
Here $b_{p,i}$ is the one-loop $\beta$-function of a 4D coupling at fixed
point $\zeta_p$, which describes the logarithmic running due to bulk zero-modes
with nonvanishing wave function at $p$. This logarithmic contribution
is relevant for the `differential running', i.e., the difference
between contributions to different brane kinetic terms \cite{Nomura:2001mf,
Hebecker:2001wq}. In general, contributions of higher Kaluza-Klein (KK) modes can
also lead to power corrections in the differential running. However,
in the considered model, such terms do not appear \cite{Hall:2001xr}.

At the fixed point $\zeta_{\rm PS}$, with unbroken subgroup $SU(4)\times
SU(2)_{\rm  L} \times SU(2)_{\rm R}$, the bulk fields yield the following
chiral superfields with positive parity \cite{Buchmuller:2017vut}: $\psi,\chi,\Psi: (4^*,1,2)$;
$\psi^c,\chi^c,\Psi^c: (4,1,2)$; $H_1,H_2: (1,2,2)$; $H_3,\ldots
H_8: (6,1,1)$, and the $\mathcal{N} = 1$ vector multiplets contained in $45: (15,1,1) + (1,3,1) + (1,1,3)$.
The three brane kinetic terms are given by
\begin{equation}\label{PSbrane}
G^{\rm PS}_1 = \frac{b_4}{8\pi^2} \ln\left(\frac{\Lambda}{M_c}\right)\ ,\quad
G^{\rm PS}_2 = \frac{b_{2L}}{8\pi^2} \ln\left(\frac{\Lambda}{M_c}\right)\ ,\quad
G^{\rm PS}_3 = \frac{b_{2R}}{8\pi^2} \ln\left(\frac{\Lambda}{M_c}\right)\ .
\end{equation}
Inserting the indices for the above representations
\cite{Slansky:1981yr}  into the
expression for the one-loop $\beta$-function, one finds $b_4 = 0$,
$b_{2L} = -4$ and $b_{2R} = 8$. 

Analogously, at the fixed point $\zeta_{\rm fl}$, with unbroken
subgroup $SU(5)'\times U(1)'$, the chiral superfields with positive
parity are \cite{Buchmuller:2017vut}: $\psi: 5^{'*}_3 +  1'_{-5}$; $\chi,\Psi: 10'_{-1}$; $\psi^c:
5'_{-3}, 1'_{5}$; $\chi^c,\Psi^c: 10^*_1$; $H_1\ldots H_8: 4\times
5^{'*}_{-2}, 4\times 5'_2$. For the $\mathcal{N} = 1$ vector multiplet
one has $45: 24'_0, 1'_0$. The two brane kinetic terms read
\begin{equation}\label{flbrane}
G^{\rm fl}_1 = \frac{b_{5'}}{8\pi^2} \ln\left(\frac{\Lambda}{M_c}\right)\ ,\quad
G^{\rm fl}_2 = \frac{b_{1'}}{8\pi^2} \ln\left(\frac{\Lambda}{M_c}\right)\ .
\end{equation}
Inserting the indices of the $SU(5)'$ representations \cite{Slansky:1981yr} into the one-loop
$\beta$-function, one obtains $b_{5'} = -4$ and $b_{1'} = 17/8$.

Eqs.~\eqref{PSbrane} and \eqref{flbrane}, together with
Eqs.~\eqref{gauge_uni} and $\Lambda/M_c \simeq 3.7$, yield the result
\begin{equation}
\delta_{23} = -0.8\ , \quad \delta_{12} = 3.1\ .
\end{equation}
Inspection of  Fig.~\ref{fig2} and Fig.~\ref{fig3} shows that this
result is incompatible with the values of $\delta_{23}$ and $\delta_{12}$ required by gauge coupling
unification. Hence, given just the bulk field content of the
considered model, unification of gauge couplings cannot be achieved at
strong coupling.  However, the differential running beyond the
compactification scale is strongly model dependent. Adding further
heavy vector-like pairs of bulk fields and/or $\mathcal{N}=1$ split multiplets at the branes does not
change the low-energy phenomenology, but it
significantly modifies the values of $\delta_{23}$ and
$\delta_{12}$. Indeed, orbifold compactifications of the heterotic
string generically yield as many $\mathcal{N}=1$ split multiplets on the branes as bulk fields. 
Adding on the PS brane $N_4$ pairs of $(4,2,1), (4^*,2,1)$,
$\bar{N}_4$ pairs of $(4,1,2), (4^*,1,2)$, $N_6$ copies of $(6,1,1)$
and $N_2$ copies of $(1,2,2)$, and on the flipped brane $N_{5'}$ pairs
of $5'_2,5^{'*}_{-2}$, $N_{5',1'}$ pairs of
$5^{'*}_3,1'_{-5},5'_{-3},1'_5$, and $N_{10'}$ pairs of $10'_{-1},
10^{'*}_1$, one obtains
\begin{equation}
\begin{split}
\delta_{23} = &\frac{1}{2\pi}\left(-4 + 2N_4 - 2\bar{N}_4 - N_6 +
  N_2\right) \ln\left(\frac{\Lambda}{M_c}\right)\ , \\
\delta_{12} = &\frac{1}{2\pi}\left(\frac{44}{5} - \frac{2}{5}\left(2N_4 - 2\bar{N}_4 - N_6 +
  N_2\right)\right. \\
& \left. \qquad+ \frac{3}{25}\left(49-23 N_{10'} - 6 N_{5'} - N_{5',1'}\right)\right)
%
\ln\left(\frac{\Lambda}{M_c}\right)\ .
\end{split}
\end{equation}
As an example, the choice $N_4=N_2=4$, $\bar{N}_4 = N_{5'} = N_{5',1'} = 3$,
$N_6=1$, $N_{10'}=12$, with $\Lambda/M_c \simeq 3.7$, yields 
$\delta_{23} = 0.2$, $\delta_{12} = -5.3$, in agreement with gauge
coupling unification for case~A, the THDM without light higgsinos. One
can also find examples that give gauge coupling unification in
case~B. However, we do not want to emphasize this possibility since
the considered model is incomplete in any case. For the bulk field content
of our model the irreducible $SO(10)$ 6D anomalies are
satisfied. But one also has to satisfy irreducible and reducible
gravitational and $U(1)_A$ anomalies as well as fixed point anomalies
\cite{Asaka:2002my,vonGersdorff:2002us}, which requires further bulk and brane fields. Hence, in the
considered model, we cannot compute the compactification scale, we can
only demonstrate that phenomenologically acceptable extensions of the
model can be consistent with gauge coupling unification and proton
decay.

The required brane kinetic terms are rather large, corresponding to
corrections of $\alpha_i^{-1}$ up to 10\%. Because of the large number
of bulk and brane fields the perturbative treatment of the model
breaks down close to the compactification scale $M_c \simeq 2\times
10^{15}~\mathrm{GeV}$. We conclude that starting from the considered
model, the compactification scale cannot be consistently increased
much by adding further vector-like fields. Hence, our model is only
consistent for a proton lifetime close to the current lower bound.

\section{Proton decay}
\label{sec:decay}

For the considered 6D $SO(10)$ model proton decay rates can be
evaluated in the standard manner. The model determines the currents
that couple to vector bosons carrying $B-L$ charge. Knowing
the zero-mode wave functions of quarks and leptons,  their  couplings to
the various vector boson KK-modes can be computed. Integrating out the
vector boson KK-modes provides the dimension-six operators whose
matrix elements determine the proton decay rates. As we shall see, a
special feature of our model is the flavour structure of these
operators, which leads to unexpected branching fractions.

\subsection{Effective operators and decay widths}
The interaction Lagrangian of the 16-plets with $SO(10)$ gauge fields is given by
\be \label{gauge}
\mathcal{L}_I = \int d^4\theta\, \left( \overline{\psi}\, e^{2 g_{\rm 6D}
    V}\, \psi +  \overline{\psi^c}\, e^{-2 g_{\rm 6D} V}\, \psi^c +
  \overline{\chi}\, e^{2 g_{\rm 6D} V}\, \chi+  \overline{\chi^c}\, e^{-2 g_{\rm 6D}
    V}\, \chi^c \right) \ ,
\ee
where $V$ is a vector field in the
adjoint representation of $SO(10)$. The decomposition of the 45-plet
with respect to $SU(3)_C\times SU(2)_L \times U(1)_Y \times U(1)_X$
is listed in Table~\ref{tab_45plet}. 
\begin{table}[t]
\begin{center}
\begin{tabular}{ccc} 
\hline
\hline
\\[-1em]
~~~$SU(5)\times U(1)_X$~~~ & ~~~$SU(3)_C\times SU(2)_L \times U(1)_Y \times U(1)_X$~~~& ~~~$U(1)_{B-L}$~~~\\
\\[-1em]
\hline
\\[-1em]
$(24,0)$ & $(8,1,0,0)$ &  0\\ 
\\[-1em]
  & $(3,2,-\frac{5}{6},0) + (\overline{3},2,\frac{5}{6},0)$ & $\mp \frac{2}{3}$ \\ 
  \\[-1em]
  & $(1,3,0,0)$ & 0\\ 
  \\[-1em]
  & $(1,1,0,0)$ & 0\\ 
  \\[-1em]
\hline
\\[-1em]
$(10,4)+(\overline{10},-4)$ & $(3,2,\frac{1}{6},4) +(\overline{3},2,-\frac{1}{6},-4) $ & $\mp \frac{2}{3}$ \\ 
\\[-1em]
  & $(\overline{3},1,-\frac{2}{3},4) + (3,1,\frac{2}{3},-4)$ & $\mp \frac{4}{3}$ \\ 
  \\[-1em]
  & $(1,1,1,4) + (1,1,-1,-4)$ & 0\\ 
  \\[-1em]
\hline
\\[-1em]
$(1,0)$ & $(1,1,0,0)$ & 0\\
\\[-1em]
\hline
\hline
\end{tabular}
\end{center}
\caption{The decomposition of ${\bf 45}$-plet of $SO(10)$.}
\label{tab_45plet}
\end{table}
Proton decay is induced by the exchange of the gauge bosons $X \sim (3,2,-\frac{5}{6})$,  $Y \sim (3,2,\frac{1}{6})$ and  $Z \sim (\overline{3},1,-\frac{2}{3})$, which carry nonzero $B-L$ charge. The terms in Eq.~\eqref{gauge}, involving these gauge bosons and massless fermions, are described by the 4D effective Lagrangian
\begin{align} \label{PD_lagrangian}
{\cal L}_{\rm 4D} =-i \frac{g_{\rm 4D}}{\sqrt{2}} \sum_{m,n \ge 0} &\left[ \overline{X}_{a \mu}^{(2m,2n+1)} {\cal X}^\mu_{a}(m,n)  +  \overline{Y}_{a \mu}^{(2m+1,2n+1)} {\cal Y}^\mu_{a}(m,n) \right. \nonumber\\
&+\left.  \overline{Z}_{a \mu}^{(2m+1,2n)} {\cal Z}^\mu_{a}(m,n)
\right]+{\rm h.c.}\,, 
\end{align}
where $g_{\rm 4D}=g_{\rm 6D}/\sqrt{V_2}$ is the 4D gauge coupling.
The three currents relevant for proton decay are given by
\begin{align} 
{\cal X}^\mu_{a}(m,n)  = \sum_{\alpha i}\, \big[
  &I^{(1)}_{i\alpha}(m,n)\, \epsilon_{abc}\,  \overline{q}_{b i}\,
  \gamma^\mu\, u^c_{c\alpha} +  I^{(1)*}_{i\alpha}(m,n)\,
  \overline{e^c}_{\alpha}\, \gamma^\mu\, q_{a i} \nonumber \\
 - &I^{(2)}_{i\alpha}(m,n)\, \overline{d^c}_{a\alpha}\,
  \gamma^\mu\, l_{i}\big]\,,  \label{current_X}  \\
\nonumber\\
{\cal Y}^\mu_{a}(m,n)  = \sum_{\alpha,i}\, \big[ &I^{(3)}_{i\alpha}(m,n)\,
  \epsilon_{abc}\,  \overline{q}_{b i}\, \gamma^\mu\, d^c_{c \alpha} +
  I^{(3)*}_{i\alpha}(m,n)\, \overline{n^c}_{\alpha}\, \gamma^\mu\, q_{a i}  \, \nonumber \\
 - &I^{(4)}_{i\alpha}(m,n)\, \overline{u^c}_{a \alpha}\,
   \gamma^\mu\, l_{i}\big] \,, \label{current_Y}\\
\nonumber\\
{\cal Z}^\mu_{a}(m,n)  = \sum_{\alpha,\beta}\, \big[& I^{(5)}_{\alpha\beta}(m,n)\, \overline{e^c}_{\alpha}\, \gamma^\mu\, d^c_{a \beta}\, +  I^{(5)*}_{\alpha\beta}(m,n)\, \overline{n^c}_{\beta}\, \gamma^\mu\, u^c_{a \alpha}\,\big] \nonumber \\
 -  \sum_{i,j}\, &I^{(6)}_{ij}(m,n)\, \overline{q}_{a i}\,\gamma^\mu\, l_{j}\, .\label{current_Z}
\end{align}       
Here $a,b,c=1,2,3$ are color indices, $i,j=1,2,3$ and
$\alpha,\beta=1,...,4$ are flavour indices. $m,n = 0,1,2,...$
label the KK-modes of the $X$-, $Y$- and $Z$-bosons. The dimensionless coefficients $I^{(r)}(m,n)$  ($r=1,...,6$)
arise from overlaps between the profiles of KK-modes of gauge bosons and zero-modes of fermions. Their explicit expressions are given in Appendix~\ref{app_A}. 

We observe that the currents coupled to the KK-modes of the
$X$-, $Y$- and $Z$-bosons  do not conserve flavour, i.e., one has
$I_{i \alpha}^{(r)}(m,n) \neq 0$ for $i\neq \alpha$ in
Eqs.~\eqref{current_X}, \eqref{current_Y} and \eqref{current_Z}. This
is in contrast to 4D GUTs where the couplings to $X$, $Y$ and $Z$ are
flavour diagonal \cite{Nath:2006ut,Langacker:1980js}.  The origin of
this effect can be attributed to the fact that the three flavours of
the SM quarks and leptons arise from the zero-modes of two bulk
16-plets in our model. As we will show at the end of this section,
this flavour non-diagonal current, together with flavour non-diagonal
mass matrices, gives rise to a proton decay pattern which is
qualitatively different from that predicted in the 4D GUT models.

After integrating out the KK tower of $X$-, $Y$- and $Z$-bosons, one
obtains from Eq.~(\ref{PD_lagrangian}) the following effective
operators for the proton decay:
\begin{align} \label{eff_op}
-{\cal L}^{\rm eff}_{\rm 4D} = \frac{g_{\rm 4D}^2}{2}
&\sum_{\alpha,\beta}\,  \sum_{i,j}\, \epsilon_{abc} \big[C^{(1)}_{i\alpha j\beta}\,
  \overline{u^c}_{c\beta} \gamma^\mu q_{bj}\,\,
  \overline{e^c}_{\alpha} \gamma_\mu q_{a i} 
-  C^{(2)}_{i\alpha j\beta}\,\, 
\overline{u^c}_{c\beta} \gamma^\mu q_{b j}\,\, \overline{d^c}_{a\alpha} \gamma_\mu l_{i} \nonumber\\
 &+  C^{(3)}_{i\alpha j\beta}\,\, \overline{d^c}_{c\beta} \gamma^\mu
q_{bj}\,\, \overline{n^c}_{\alpha} \gamma_\mu q_{a i} 
 -  C^{(4)}_{i\alpha j\beta}\,\, 
  \overline{d^c}_{c\beta} \gamma^\mu q_{b j}\,\,
  \overline{u^c}_{a\alpha} \gamma_\mu l_{i} \big] + {\rm h.c.\,,} 
\end{align}
with the coefficients
\begin{equation} \label{C_i}
\begin{split}
C^{(1)}_{i \alpha j \beta} &= \sum_{m,n} \frac{1}{M_X^2(m,n)}\, I^{(1)*}_{j\beta}(m,n)\, I^{(1)*}_{i\alpha}(m,n)\,, \\
C^{(2)}_{i\alpha j\beta} &= \sum_{m,n} \frac{1}{M_X^2(m,n)}\, I^{(1)*}_{j\beta}(m,n)\, I^{(2)}_{i\alpha}(m,n)\,, \\ 
C^{(3)}_{i\alpha j\beta} &= \sum_{m,n} \frac{1}{M_Y^2(m,n)}\, I^{(3)*}_{j\beta}(m,n)\, I^{(3)*}_{i\alpha}(m,n)\,, \\
C^{(4)}_{i\alpha j\beta} &= \sum_{m,n} \frac{1}{M_Y^2(m,n)}\,
I^{(3)*}_{j\beta}(m,n)\, I^{(4)}_{i\alpha}(m,n)\,, 
\end{split}
\end{equation}
where the vector boson masses are given by\footnote{Note that our
  convention for $R_1,R_2$ differs by a factor $2\pi$ from the
  convention in\cite{Buchmuller:2004eg}.}
\begin{equation} \label{XYmasses}
\begin{split}
M_X^2(m,n) &= 4 \pi^2 \left( \frac{(2m)^2}{R_1^2} + \frac{(2n+1)^2}{R_2^2}\right)\,, \\
M_Y^2(m,n) &= 4 \pi^2 \left( \frac{(2m+1)^2}{R_1^2} +
  \frac{(2n+1)^2}{R_2^2}\right)\,. 
\end{split}
\end{equation}
The $Z$-bosons by themselves do not lead to dimension-six operators
that induce proton decay. Their contribution only arises after electroweak symmetry breaking through their mixing with some $Y$-bosons. We do not include these contributions in the list of operators given in Eq.~\eqref{eff_op}. Moreover, the third term in Eq.~\eqref{eff_op} is irrelevant for proton decay as it involves heavy singlet neutrinos. Using Fierz reordering, the remaining terms can be rewritten as
\begin{equation} \label{eff_op_2}
\begin{split}
{\cal L}^{\rm eff}_{\rm 4D} = g_{\rm 4D}^2 \sum_{\alpha,\beta}\,
\sum_{i,j}\, &\big[C^{(11)}_{i\alpha j\beta}\, \epsilon_{abc}\,
\overline{e^c}_{\alpha} \overline{u^c}_{a\beta} u_{b j} d_{c i}\, \\
& + C^{(24)}_{i\alpha j\beta}\, \epsilon_{abc}\,
  \overline{d^c}_{a\alpha}  \overline{u^c}_{b\beta} \left(
    d_{c j} \nu_{i}\,  -\,   u_{c j} e_{i} \right) \big] + {\rm h.c.\,,} 
\end{split}
\end{equation}
with 
\begin{equation} \label{C'}
\begin{split}
C^{(11)}_{i\alpha j\beta} &= C^{(1)}_{i \alpha j\beta} + C^{(1)}_{j
  \alpha i \beta}\,, \\
C^{(24)}_{i\alpha j\beta} &= C^{(2)}_{i \alpha j \beta} - C^{(4)}_{i \beta j \alpha}\,. 
\end{split}
\end{equation}

The operators in the physical basis are obtained using the unitary transformations between weak ($f$) and mass ($f'$) eigenstates,
\be  \label{U}
f = U_{f}\, f'\,, \ee
where $f=u,d,e,\nu,u^c,d^c,e^c$. In the present framework, the unitary
matrices $U_{u^c}$, $U_{d^c}$ and $U_{e^c}$ are of dimension $4 \times
4$, while $U_{u}$, $U_{d}$, $U_e$ and $U_{\nu}$ are $3 \times 3$
matrices. Changing from weak to mass eigenstates,
Eq.~(\ref{eff_op_2}) becomes
\begin{equation} \label{eff_op_3}
\begin{split}
{\cal L}^{\rm eff}_{\rm 4D} = g_{\rm 4D}^2 \sum_{\gamma,\delta}\,
\sum_{l,k}\, &\big[d^{(1)}_{l\gamma k\delta}\,\epsilon_{abc}\,
\overline{e^c}'_{\gamma} \overline{u^c}'_{a\delta} u'_{b k} d'_{c l}\, \\
& +  d^{(2)}_{l\gamma k\delta}\, \epsilon_{abc}\,
\overline{d^c}'_{a\gamma}  \overline{u^c}'_{b\delta}  u'_{c k} e'_{l} 
+   d^{(3)}_{l\gamma k\delta}\, \epsilon_{abc}\,
  \overline{d^c}'_{a\gamma}  \overline{u^c}'_{b\delta} d'_{c k}
  \nu'_{l}\, \big] + {\rm h.c.\,,}  
\end{split}
\end{equation}
where
\begin{equation} \label{d}
\begin{split}
d^{(1)}_{l\gamma k\delta} &= \sum_{\alpha,\beta}\,  \sum_{i,j}\,
C^{(11)}_{i\alpha j\beta}\, \left(U_{e^c}^* \right)_{\alpha \gamma}\, \left(U_{u^c}^* \right)_{\beta \delta}\, \left(U_u \right)_{jk}\,  \left(U_d \right)_{il}\,, \\
d^{(2)}_{l\gamma k\delta} &= - \sum_{\alpha,\beta}\,  \sum_{ij}\,
C^{(24)}_{i\alpha j\beta}\, \left(U_{d^c}^* \right)_{\alpha\gamma}\, \left(U_{u^c}^* \right)_{\beta\delta}\, \left(U_u \right)_{jk}\,  \left(U_e \right)_{il}\,, \\
d^{(3)}_{l\gamma k\delta} &=  \sum_{\alpha,\beta}\,  \sum_{ij}\,
C^{(24)}_{i\alpha j\beta}\, \left(U_{d^c}^* \right)_{\alpha
  \gamma}\, \left(U_{u^c}^* \right)_{\beta \delta}\, \left(U_d
\right)_{jk}\,  \left(U_\nu \right)_{il}\,. 
\end{split}
\end{equation}
Note that the matrix $U_\nu$ drops out in proton decay branching
fractions since we sum over neutrino flavours in the final state.

The partial widths of the various proton decay channels can be evaluated
based on Eq.~\eqref{eff_op_3}. All operators 
conserve $B-L$, and therefore the proton decays into an
anti-lepton and a meson. The relevant hadronic matrix elements between
proton and meson states is obtained using chiral perturbation theory
\cite{Claudson:1981gh,Chadha:1983sj}. The resulting partial decay widths read \cite{Aoki:1999tw}:
\begin{align} \label{decay_width}
\Gamma[p \to e_i^+\pi^0] &= \frac{(m_p^2 - m_{\pi^0}^2)^2}{32\, \pi\, m_p^3 f_\pi^2}\, \alpha^2 A^2 g_{\rm 4D}^4 \left( \frac{1+D+F}{\sqrt{2}}\right)^2 \left(\left| d^{(1)}_{1i11} \right|^2+ \left| d^{(2)}_{i111}\right|^2\right), \nonumber\\
\Gamma[p \to \overline{\nu}\pi^+] &= \frac{(m_p^2 - m_{\pi^\pm}^2)^2}{32\, \pi\, m_p^3 f_\pi^2}\, \alpha^2 A^2 g_{\rm 4D}^4 \left(1+D+F \right)^2 \sum_{i=1}^3 \left| d^{(3)}_{i111} \right|^2\,, \nonumber\\
\Gamma[p \to e_i^+K^0] &= \frac{(m_p^2 - m_{K^0}^2)^2}{32\, \pi\, m_p^3 f_\pi^2}\, \alpha^2 A^2 g_{\rm 4D}^4 \left(1+ (D-F)\frac{m_p}{m_B}\right)^2 \left( \left|d^{(1)}_{2i11} \right|^2+\left| d^{(2)}_{i211}\right|^2\right), \nonumber\\
\Gamma[p \to \overline{\nu}K^+] &= \frac{(m_p^2 - m_{K^\pm}^2)^2}{32\, \pi\, m_p^3 f_\pi^2}\, \alpha^2 A^2 g_{\rm 4D}^4 \sum_{i=1}^3 \left| \frac{2D}{3}\frac{m_p}{m_B} d^{(3)}_{i211}+\left( 1+\frac{D+3F}{3}\frac{m_p}{m_B}\right) d^{(3)}_{i121}\right|^2, \nonumber\\
\Gamma[p \to e_i^+\eta] &= \frac{(m_p^2 - m_{\eta}^2)^2}{32\, \pi\, m_p^3 f_\pi^2}\, \alpha^2 A^2 g_{\rm 4D}^4 \left(\frac{1+D-3F}{\sqrt{6}}\right)^2 \left(\left| d^{(1)}_{1i11} \right|^2+ \left| d^{(2)}_{i111} \right|^2\right).
\end{align}
Here $e_i^+ = (e^+, \mu^+)$, $m_h$
($h=p,\pi^0,\pi^\pm,K^0,K^\pm,\eta$) denotes the mass of hadron $h$,
$f_\pi$ is the pion decay constant and $m_B$ is the average baryon
mass. The factors 
$\alpha$, $D$ and $F$ are parameters of the chiral Lagrangian while
$A$ incorporates renormalization group running effects of the hadronic
matrix elements.

\subsection{Numerical results}
We now evaluate the partial proton decay widths for the two fits of the
flavour spectrum performed in \cite{Buchmuller:2017vut}. In the first
one (Fit~I), the Yukawa couplings and brane mass parameters of the
model were determined by fitting fermion masses and mixing
parameters using the $\chi^2$ minimization method. We obtained a very
good fit with a minimum $\chi^2 = 0.5$; in this case, leptogenesis led to a
baryon asymmetry two orders of magnitude below the observed value. We
then performed another fit (Fit~II) where the observed baryon asymmetry was
used as a constraint, for which we obtained a
minimum $\chi^2 = 0.95$. For both fits, we give the
various unitary matrices $U_f$ that connect weak and mass eigenstates 
in Appendix~\ref{U_extraction}. 

The effective couplings $C_i$ in Eq.~(\ref{C_i}) involve summations
over the KK-modes of gauge bosons.  We notice that the value of some
of the $C_i$'s decrease slightly  when contributions from higher
KK-modes are taken into account because of destructive interference
between different amplitudes. We also find that the values of these
coefficients converge rapidly, and only the contributions from the first few KK-modes are relevant. We consider the first five modes, corresponding to $m,n=0,1,...,4$, for the evaluation of the partial widths. 

We use the parameters $\alpha = 0.01$ GeV$^{3}$, $D=0.8$ and $F=0.46$
for the chiral Lagrangian  parameters \cite{Aoki:1999tw}. The
parameter $A$ can be written as $A=A_{\rm SD} A_{\rm LD}$, where
$A_{\rm LD}$ takes into account the renormalization effects from $M_Z$
to the proton mass scale while $A_{\rm SD}$ includes the short
distance running effects from $M_{\rm GUT}$ to $M_Z$.  For our
calculation, we use $A_{\rm LD} =1.43$ and $A_{\rm SD} =2.26$
\cite{Alonso:2014zka}. The values of hadron masses are taken from the
PDG \cite{Tanabashi:2018oca}. We use $m_B = 1.15$ GeV as average
baryon mass, the pion decay constant $f_\pi = 130$ MeV, and  we take
$g_{\rm 4D} = 0.57$. The compactification scale is identified with the
mass of the lightest KK-mode of the $X$-, $Y$-bosons, which
corresponds to $R_1=R_2=2 \pi M_c^{-1}$.

The proton lifetime in a particular channel, $p \to \bar{l} + X$, is defined as
\be \label{tau}
\tau/{\rm BR}[p\to \bar{l} + X] = 1/\Gamma[p\to \bar{l} + X]\,,
\ee
where $\tau = 1/\Gamma_{\rm total}$, and the branching ratio is ${\rm
  BR}[p\to \bar{l} + X] = \Gamma[p\to \bar{l} + X] / \Gamma_{\rm total}$. The
current limits on the proton lifetime are at $90\% $ confidence level \cite{Miura:2016krn},
\begin{equation} \label{exp_bounds}
\tau/{\rm BR}[p\to e^+ \pi^0] > 1.6\times 10^{34}\,{\rm ~yrs}, \quad
\tau/{\rm BR}[p\to \mu^+ \pi^0]  > 7.7\times 10^{33}\,{\rm ~yrs}\,. 
\end{equation}
With the aforementioned values of the various parameters, we obtain
for Fit~I,
\beqa \label{tau_fit1}
\tau/{\rm BR}[p\to e^+ \pi^0] & = & 1.6\times 10^{34}\,{\rm
  ~yrs} \times \left(\frac{M_c}{2.2 \times 10^{15}\,{\rm GeV}}\right)^4\,, \nonumber \\
\tau/{\rm BR}[p\to \mu^+ \pi^0] & = & 4.7\times 10^{34}\,{\rm ~yrs} \times \left(\frac{M_c}{2.2 \times 10^{15}\,{\rm GeV}}\right)^4\,, \eeqa
and for Fit~II,
\beqa \label{tau_fit2}
\tau/{\rm BR}[p\to e^+ \pi^0] & = & 1.6\times 10^{34}\,{\rm
  ~yrs} \times \left(\frac{M_c}{2.0 \times 10^{15}\,{\rm GeV}}\right)^4\,, \nonumber \\
\tau/{\rm BR}[p\to \mu^+ \pi^0] & = & 7.7\times 10^{33}\,{\rm ~yrs} \times \left(\frac{M_c}{2.0 \times 10^{15}\,{\rm GeV}}\right)^4\,, \eeqa
In case of the latter, the channel $p\to \mu^+ \pi^0$
provides an equally strong constraint as the positron channel. 
These numbers have to be compared with $\tau/{\rm BR}[p\to e^+ \pi^0] \sim 2\times 10^{35}{\rm
  ~yrs}$,  the experimental reach of Hyper-Kamiokande
\cite{Abe:2018uyc}, and with 
$\tau/{\rm BR}[p\to e^+ \pi^0] \sim {\rm few} \times 10^{34}{\rm
  ~yrs}$, the experimental reach of DUNE \cite{Acciarri:2015uup}.

The various branching fractions obtained for Fit~I and Fit~II are displayed in Table \ref{tab:pd}.
\begin{table}[t]
\begin{center}
\begin{tabular}{lrr} 
\hline
\hline
Branching ratio [\%]& ~~~~~~Fit~I  & ~~~~~~Fit~II\\
\hline
${\rm BR}[p\to e^+ \pi^0]$ & 46 & 23\\
${\rm BR}[p\to \mu^+ \pi^0]$  & 17 & 51\\
${\rm BR}[p\to \bar{\nu} \pi^+]$  & 5 &3\\
${\rm BR}[p\to e^+ K^0]$ & 17 & 15\\
${\rm BR}[p\to \mu^+ K^0 ]$  & 12 & 6\\
${\rm BR}[p\to \bar{\nu} K^+]$  & 1 & 1\\
${\rm BR}[p\to e^+ \eta]$ & $<$ 1 & $<$ 1\\
${\rm BR}[p\to \mu^+ \eta]$  & $<$ 1 & $<$ 1\\
\hline
\end{tabular}
\end{center}
\caption{Proton decay branching fractions for two different fits to
  the flavour spectrum.}
\label{tab:pd}
\end{table}
A noteworthy feature of these results is that the branching ratios of
proton decay into the channels  involving $e^+$ or $\mu^+$ are of the
similar magnitude. The origin of this in the present model can be
understood in the following way. First, as already mentioned earlier,
the massless modes of quark and lepton generations have flavour
non-diagonal overlaps with the KK-modes of heavy vector bosons. In
particular, the flavours arising from the 
zero-modes of $\psi$ have comparable diagonal and non-diagonal
couplings with the $X$- and $Y$-bosons. For example, zero-mode overlap
integrals $I^{(1)}_{i\alpha}(0,0)$ ($I^{(2)}_{i \alpha}(0,0)$) of
fermion zero-modes and vector bosons with $m=n=0$, that cause proton
decay into a charged lepton with flavour $\alpha$ ($i$) and a quark
with flavour $i$ ($\alpha$), are given by
\beqa \label{integrals}
\left(I^{(1)}_{i\alpha}(0,0)\right) & =&\left(
\begin{array}{cccc}
 -0.12 & -0.07 & -1.0 & 0 \\
 0.12 & 1.0 & 0.07 & 0 \\
 0 & 0 & 0 & 0.45 \\
\end{array}
\right)\,, \nonumber \\
\left(I^{(2)}_{i\alpha}(0,0)\right) & =&\left(
\begin{array}{cccc}
 -0.28 & -0.04 (1+ i) & 0 & 0 \\
 0.1 (1 - i) & 1.0 & 0 & 0 \\
 0 & 0 & 0.51 & -0.51 \\
\end{array}
\right)\,, \eeqa
These overlap integrals are properties of the wave functions of
fermions and gauge bosons, which are unambiguously determined in the
considered model and which do not depend on the Yukawa structure,
i.e., on the particular fit of the flavour structure.

Second, as shown in Appendix \ref{U_extraction}, some of the
unitary matrices introduce large flavour mixings. In particular, the
matrices $U_{u^c}$, $U_{d^c}$ and $U_{e^c}$ give rise to a strong
mixing between the first and second generation of $u^c$, $d^c$ and
$e^c$, respectively, both in Fit~I and in Fit~II. This feature depends
on the Yukawa structure of the theory and hence the precise results
are fit dependent. In the present framework, the non-diagonal gauge
boson couplings as well as the flavour structure of the mass matrices
give rise to an enhanced branching fraction ${\rm BR}[p\to \mu^+ \pi^0]$, comparable to the branching fraction 
${\rm BR}[p\to e^+ \pi^0]$. This prediction is different from typical
predictions made in the context of 4D GUTs, and it also distinguishes
the 6D $SO(10)$ model with magnetic flux from the $SO(10)$ model
without magnetic flux \cite{Buchmuller:2004eg}.

\section{Summary and conclusions}
\label{sec:conclusion}

In flux compactifications the quark-lepton generations are zero-modes
with characteristic wave functions in the compact space. In the
considered orbifold GUT model the Yukawa matrices are determined by
the values of these wave functions at the orbifold fixed points,
together with complex couplings of bulk fields at the fixed points
which are generation independent. At each fixed point this leads to
rank-one Yukawa matrices with $\mathcal{O}(1)$ entries.
In addition there are large mass mixing terms with charge-conjugate
split multiplets. The resulting up-quark, down-quark, charged lepton and
Dirac neutrino mass matrices have large off-diagonal entries, and the
same is true for the unitary matrices which diagonalize them. For two
fits of the flavour spectrum, these matrices are given in Appendix~B.
A small mismatch of these matrices for up-quarks and down-quarks leads
to small off-diagonal terms in the CKM matrix, whereas large off-diagonal
terms appear in the PMNS matrix due to the seesaw mechanism. 

As a consequence, the unitary matrices that connect weak and mass
eigenstates lead to large flavour non-diagonal couplings of vector
bosons with non-zero $B-L$ charge, whose exchange induces proton decay. This
effect is enhanced by non-diagonal couplings of these heavy vector bosons to
flavour eigenstates, which result from overlap integrals of vector
boson and fermion mode functions. This is in contrast to 4D GUT models
and other orbifold GUT models without flux, where the couplings of heavy vector bosons to flavour
eigenstates are diagonal. 

In the considered 6D $SO(10)$ orbifold GUT model gauge coupling
unification can not be achieved at some cut-off scale beyond the
compactification scale. Hence, the compactification scale cannot be
computed. Gauge coupling unification requires the addition of further
vector-like bulk and/or brane fields which would not affect the
low-energy phenomenology. Due to the large number of vector-like
fields the theory becomes strongly interacting already close to the
compactification scale. It appears that matching the
non-supersymmetric Standard Model to a higher-dimensional
supersymmetric theory requires a compactification scale that is barely
consistent with current constraints from proton decay.
An unexpected prediction of our model is the similar size
of the branching ratios $p \rightarrow e^+\pi^0$ and $p\rightarrow
\mu^+\pi^0$ , where the latter one can even be
dominant. This appears to be a generic feature of flux
compactifications of higher-dimensional GUT models.

\section*{Acknowledgements}
We thank Emilian Dudas, Arthur Hebecker, Yoshiyuki Tatsuta and
Alexander Westphal for valuable discussions. This work was supported by the German Science Foundation (DFG) within the Collaborative Research Center (SFB) 676 ``Particles,
Strings and the Early Universe''. The work of KMP  was partially supported by a research grant under INSPIRE Faculty Award (DST/INSPIRE/04/2015/000508) from the Department of Science and Technology, Government of India. KMP thanks the DESY Theory Group for the kind hospitality during the initial stage of this work.

\appendix
\begin{appendix}
\section{Definition of overlap integrals}
\label{app_A}
The dimensionless overlap integrals introduced in Section~\ref{sec:decay} are
given as
\begin{equation} \label{I1}
\begin{split}
I^{(1)}_{i' \alpha'}(m,n)  &= \sqrt{V_2} \int d^2y\, \psi_{-+}^{(i') *}(y)\, \psi_{++}^{(\alpha')}(y)\, f_X^{(m,n)}(y)\,, \\
I^{(1)}_{34}(m,n)  &= \sqrt{V_2}\int d^2y\, \chi_{--}^{(1) *}(y)\, \chi_{+-}^{(1)}(y)\, f_X^{(m,n)}(y)\,, \\
I^{(1)}_{i'4}(m,n) &=  I^{(1)}_{3\alpha'}(m,n) =0\,.
\end{split}
\end{equation}

\begin{equation} \label{I2}
\begin{split}
I^{(2)}_{i'j'}(m,n) &= \sqrt{V_2}\int d^2y\, \psi_{+-}^{(j')*}(y)\, \psi_{--}^{(i')}(y)\, f_X^{(m,n)}(y)\,, \\
I^{(2)}_{3(j'+2)}(m,n) &= \sqrt{V_2}\int d^2y\, \chi_{++}^{(j')*}(y)\, \chi_{-+}^{(1)}(y)\, f_X^{(m,n)}(y)\,, \\
I^{(2)}_{i'(j'+2)}(m,n) &= I^{(2)}_{3j'}(m,n) = 0\,.
\end{split}
\end{equation}

\begin{equation} \label{I3}
\begin{split}
I^{(3)}_{i'j'}(m,n)  &= \sqrt{V_2}\int d^2y\, \psi_{-+}^{(i') *}(y)\, \psi_{+-}^{(j')}(y)\, f_Y^{(m,n)}(y)\,, \\
I^{(3)}_{3(j'+2)}(m,n)  &= \sqrt{V_2}\int d^2y\, \chi_{--}^{(1) *}(y)\, \chi_{++}^{(j')}(y)\, f_Y^{(m,n)}(y)\,, \\
I^{(3)}_{i'(j'+2)}(m,n) & =  I^{(3)}_{3,j'}(m,n)=0\,.
\end{split}
\end{equation}

\begin{equation} \label{I4}
\begin{split}
I^{(4)}_{i'\alpha'}(m,n)  &= \sqrt{V_2}\int d^2y\, \psi_{++}^{(\alpha') *}(y)\, \psi_{--}^{(i')}(y)\, f_Y^{(m,n)}(y)\,, \\
I^{(4)}_{34}(m,n)  &= \sqrt{V_2}\int d^2y\, \chi_{+-}^{(1) *}(y)\, \chi_{-+}^{(1)}(y)\, f_Y^{(m,n)}(y)\,, \\
I^{(4)}_{i'4}(m,n) &= I^{(4)}_{3\alpha'}(m,n) =0\,.
\end{split}
\end{equation}

\begin{equation} \label{I5}
\begin{split}
I^{(5)}_{\alpha'i'}(m,n)  &= \sqrt{V_2}\int d^2y\, \psi_{++}^{(\alpha') *}(y)\, \psi_{+-}^{(i')}(y)\, f_Z^{(m,n)}(y)\,, \\
I^{(5)}_{4(j'+2)}(m,n)  &= \sqrt{V_2}\int d^2y\, \chi_{+-}^{(1) *}(y)\, \chi_{++}^{(j')}(y)\, f_Z^{(m,n)}(y)\,, \\
I^{(5)}_{4j'}(m,n) &= I^{(5)}_{44}(m,n) =0\,.
\end{split}
\end{equation}

\begin{equation}\label{I6}
\begin{split}
I^{(6)}_{i'j'}(m,n) &= \sqrt{V_2}\int d^2y\, \psi_{-+}^{(i') *}(y)\, \psi_{--}^{(j')}(y)\, f_Z^{(m,n)}(y)\,, \\
I^{(6)}_{33}(m,n) &= \sqrt{V_2}\int d^2y\, \chi_{--}^{(1) *}(y)\, \chi_{-+}^{(1)}(y)\, f_Z^{(m,n)}(y)\,, \\
I^{(6)}_{i'3}(m,n) &= I^{(6)}_{3j'}(m,n) =0\, .
\end{split}
\end{equation}
Here $i',j'=1,2$ and $\alpha'=1,2,3$ are the flavour indices
introduced in the decomposition of the 16-plets in
Section~\ref{sec:model}. The integration domain is given by $y_1\in
[0,R_1/2]$ and $y_2 \in [0,R_2]$. The volume of the orbifold is $V_2 = R_1 R_2/2$.

The mode functions of the gauge bosons are obtained from
\cite{Buchmuller:2004eg}. They read:
\begin{equation} \label{profiles_XYZ}
\begin{split}
f_X^{(m,n)}(y) &=\sqrt{\frac{2}{V_2}}\,  \cos\left[2 \pi \left( 2m \frac{y_1}{R_1}+(2n+1) \frac{y_2}{R_2}\right) \right] \,, \\
f_Y^{(m,n)}(y) &= \sqrt{\frac{2}{V_2}}\,\cos\left[2 \pi \left( (2m+1)
    \frac{y_1}{R_1}+(2n+1) \frac{y_2}{R_2}\right) \right] \,, \\
f_Z^{(m,n)}(y) &= \sqrt{\frac{2}{V_2}}\, \cos\left[2 \pi \left( (2m+1)
    \frac{y_1}{R_1}+2n \frac{y_2}{R_2}\right) \right] \, .
\end{split}
\end{equation}
The wavefunction profiles of quarks and leptons are obtained from the expression  \cite{Buchmuller:2017vho,Buchmuller:2017vut}
\beqa \label{profile}
\varphi_{\eta_{\rm PS},\eta_{\rm GG}}^{(j)} (y_1,y_2;N) &=& {\cal N} \exp\left(2 \pi i N \tau \frac{y_2^2}{R_2^2} \right)\, \sum_{n \in {\mathbb{Z}}} \cos \left[ 2\pi \left(-2nN+j+\frac{k_{\rm PS}}{2}\right) \left(\frac{y_1}{R_1}+\tau \frac{y_2}{R_2}\right)\right]\nonumber \\
& \times & \exp\left( 2 \pi i \tau N \left( n-\frac{j}{2N}\right)^2 -i \pi \left( n-\frac{j}{2N}\right)(k_{\rm PS} \tau -k_{\rm GG})\right)
\eeqa
where $\eta_{\rm PS} = e^{i \pi k_{\rm PS}}$, $\eta_{\rm GG} = e^{i \pi k_{\rm GG}}$ and $k_{\rm PS},\,k_{\rm GG} = 0,1$. Here $j = 0,...,N$ for $k_{\rm PS} = k_{\rm GG} =0$ and $j=0,...,N-1$ otherwise. We identify
\begin{align} \label{profile-cov}
\psi^{(2)}_{\eta_{\rm PS},\eta_{\rm GG}}(y_1,y_2) &=  \varphi^{(0)}_{\eta_{\rm PS},\eta_{\rm GG}}(y_1,y_2;2),~ \nonumber \\
\psi^{(1)}_{\eta_{\rm PS},\eta_{\rm GG}}(y_1,y_2) &=  \varphi^{(1)}_{\eta_{\rm PS},\eta_{\rm GG}}(y_1,y_2;2),~ \nonumber \\
 \psi^{(3)}_{++}(y_1,y_2) &=  \varphi^{(2)}_{++}(y_1,y_2;2),~ \nonumber \\
 \chi^{(1)}_{\eta_{\rm PS},\eta_{\rm GG}}(y_1,y_2) &=  \varphi^{(0)}_{\eta_{\rm PS},\eta_{\rm GG}}(y_1,y_2;1),~ \nonumber \\
 \chi^{(2)}_{++}(y_1,y_2) &=  \varphi^{(1)}_{++}(y_1,y_2;1).  
\end{align}

\section{Extraction of unitary matrices}
\label{U_extraction}
The $3\times 3$ matrices $U_f$ for $f=u,d,e,\nu$ are obtained
following the same procedure as in Appendix A of \cite{Buchmuller:2017vut}. For $f=u^c,d^c,e^c$ the corresponding $4\times 4$ unitary matrices are obtained in the following way. Starting from the expressions for $M_f$ given in Eq. (A2) in \cite{Buchmuller:2017vut}, we obtain
\be \label{Hcf}
H^c_f \equiv M_f^\dagger M_f = H^c_{f4} + H^c_{f3}\,, \ee
where 
\beqa
\left(H^c_{f4}\right)_{\alpha \beta} &=& \mu^{f*}_{\alpha} \mu^{f}_{\beta}\, \nonumber \\
\left(H^c_{f3}\right)_{\alpha \beta} &=& v_f^2 \left(Y_f^\dagger Y_f\right)_{\alpha \beta}\,. \eeqa
We determine $4\times 4$ unitary matrices $V_4^f$ such that 
\be V_4^{f \dagger} H^c_{f4} V_4^f = {\rm Diag.} \left(0,0,0,\tilde{\mu}_f^2\right) \equiv D_4^f\,,\ee
where $\tilde{\mu}_f^2=\sum_{\alpha}|\mu_{\alpha}|^2$. We then obtain 
\be \label{Hcf2}
\tilde{H}^c_f = V_4^{f \dagger} H^c_f V_4^{f} =D_4^f + V_4^{f \dagger} H^c_{f3} V_4^f\,, \ee
and integrate out the heavy state to obtain an effective $3 \times 3$ matrix for the light fermions, whose elements are
\be \label{Hcf3}
\left(\tilde{H}^c_{3f}\right)_{ij} = \left(\tilde{H}^c_f\right)_{ij} - \frac{1}{\tilde{\mu}_f^2} \left(\tilde{H}^c_f\right)_{i4}\left(\tilde{H}^c_f\right)_{j4}\,. \ee
The matrix $\tilde{H}^c_{3f}$ is then diagonalized using a $3\times 3$ unitary matrix $V^f_3$ such that $V_3^{f \dagger}\tilde{H}^c_{3f} V_3^f = D_3^f$. Finally,
the $4 \times 4$ unitary matrix
\be \label{Uf}
U_f = V_4^f \left( \ba{cc} V_3^f & 0\\ 0 & 1 \ea \right) \ee
is constructed, which relates the weak eigenstates $(f)$ with mass eigenstates $(f')$ such that $f = U_f f'$, for $f=u^c,d^c,e^c$.

Following the above procedure, the unitary matrices obtained from  Fit-I read
\begin{small}
\beqa \label{Uf1}
U_u & =& \left(
\begin{array}{ccc}
 -0.9748+0.0693 i & -0.2117+0.0008 i & -0.0105 \\
 0.2114\, -0.015 i & -0.9761+0.0035 i & -0.0486+0.0002 i \\
 0 & -0.0497 & 0.9988 \\
\end{array}
\right)\,, \nonumber \\
U_d &=& \left(
\begin{array}{ccc}
 -0.1438-0.9543 i & -0.1447+0.2179 i & -0.0156+0.0003 i \\
 -0.1905+0.1788 i & -0.9611+0.013 i & -0.0887+0.0047 i \\
 -0.0198 & -0.088 & 0.9959 \\
\end{array}
\right)\,, \nonumber \\
U_e &=&\left(
\begin{array}{ccc}
 -0.0792-0.1946 i & -0.7026+0.6796 i & -0.0097-0.0152 i \\
 -0.9734+0.0162 i & -0.0817-0.1944 i & -0.0886+0.0008 i \\
 -0.0904 & -0.0036 & 0.9959 \\
\end{array}
\right)\,, \nonumber \\
U_{u^c} &=&\left(
\begin{array}{cccc}
 0.1981\, +0.6784 i & -0.4109+0.5744 i & -0.0248+0.0347 i & 0 \\
 -0.1395-0.4804 i & -0.2907+0.4053 i & -0.0184+0.0245 i & -0.7071 \\
 0.1395\, +0.4804 i & 0.2909\, -0.4053 i & 0.0167\, -0.0245 i & -0.7071 \\
 0.0001 & -0.0603 & 0.9982 & -0.0012 \\
\end{array}
\right)\,, \nonumber\\
U_{d^c} &=&\left(
\begin{array}{cccc}
 -0.949-0.1635 i & 0.2576\, -0.0511 i & 0.0061\, +0.0604 i & 0 \\
 -0.0776+0.0864 i & -0.2752+0.3064 i & -0.443+0.4931 i & 0.\, +0.6144 i \\
 0.0604\, -0.0672 i & 0.2143\, -0.2386 i & 0.3449\, -0.384 i & 0.\, +0.789 i \\
 -0.1725+0.1458 i & -0.2797+0.7617 i & 0.204\, -0.4989 i & 0 \\
\end{array}
\right)\,, \nonumber\\
U_{e^c} &=& \left(
\begin{array}{cccc}
 0.7074 & 0.7053 & -0.0473 & 0 \\
 -0.4998 & 0.4988 & -0.0373 & -0.7071 \\
 0.4998 & -0.4989 & 0.0355 & -0.7071 \\
 -0.003 & 0.0698 & 0.9976 & -0.0012 \\
\end{array}
\right)\,.
\eeqa
\end{small}

Correspondingly, 
the unitary matrices obtained from the Fit II are given by
\begin{small}
\beqa \label{Uf2}
U_u & =& \left(
\begin{array}{ccc}
 -0.5613-0.8003 i & -0.0123+0.2106 i & 0.0001\, -0.0018 i \\
 0.1219\, +0.1722 i & -0.0527+0.976 i & 0.0004\, -0.0082 i \\
 0 & 0.0084 & 1. \\
\end{array}
\right)\,, \nonumber \\
U_d &=& \left(
\begin{array}{ccc}
 -0.0205-0.9329 i & -0.3053-0.1897 i & -0.0116+0.0013 i \\
 0.2196\, +0.2847 i & -0.9294+0.0744 i & -0.0379+0.0002 i \\
 0.0092 & -0.0385 & 0.9992 \\
\end{array}
\right)\,, \nonumber \\
U_e &=& \left(
\begin{array}{ccc}
 0.1388\, +0.1353 i & -0.0003-0.0314 i & -0.6933+0.6933 i \\
 0.9743\, +0.0005 i & -0.1128+0.0044 i & 0.\, -0.195 i \\
 0.1149 & 0.9931 & 0.0225 \\
\end{array}
\right)\,, \nonumber \\
U_{u^c} &=& \left(
\begin{array}{cccc}
 0.504\, +0.4965 i & -0.4195-0.1861 i & 0.4911\, +0.2182 i & 0 \\
 -0.3553-0.3509 i & -0.8338-0.1323 i & -0.1115+0.1543 i & 0.0417 \\
 0.3563\, +0.3509 i & -0.2391+0.1323 i & -0.8061-0.1543 i & 0.0417 \\
 0 & 0.0448 & 0.0383 & 0.9983 \\
\end{array}
\right)\,, \nonumber\\
U_{d^c} &=& \left(
\begin{array}{cccc}
 -0.4785-0.848 i & 0.1196\, +0.1932 i & -0.0183+0.0012 i & 0 \\
 0.0059\, +0.0097 i & 0.0296\, +0.0343 i & -0.0553-0.0653 i & 0.\, +0.9952 i \\
 -0.0604-0.0995 i & -0.3021-0.3501 i & 0.5647\, +0.6665 i & 0.\, +0.0975 i \\
 0.1957 & 0.8558 & 0.4788 & 0 \\
\end{array}
\right)\,, \nonumber\\
U_{e^c} &=& \left(
\begin{array}{cccc}
 0.6851 & -0.4879 & 0.5409 & 0 \\
 -0.5398 & -0.8377 & -0.072 & 0.0417 \\
 0.4892 & -0.2412 & -0.8371 & 0.0417 \\
 0.0021 & 0.045 & 0.0379 & 0.9983 \\
\end{array}
\right)\,.
\eeqa
\end{small}

\end{appendix}

\providecommand{\href}[2]{#2}\begingroup\raggedright\endgroup

\end{document}